\documentclass[twocolumn,amsmath,amssymb,aps,preprintnumbers,showpacs,superscriptaddress,floatfix]{revtex4}

\usepackage{bm}         % bold math
\usepackage{amssymb}
\usepackage{amsmath}
\usepackage{graphicx}
\usepackage{epsfig}
\usepackage{color}
\usepackage{ulem}
\usepackage{SIunits}
\begin{document}

\title{Broadband electron spin resonance from $\boldsymbol{500~\mega\hertz}$ to $\boldsymbol{40~\giga\hertz}$ using superconducting coplanar waveguides}

\author{Conrad~Clauss}
\affiliation{1. Physikalisches Institut, Universit\"{a}t Stuttgart, Pfaffenwaldring 57, D-70550 Stuttgart, Germany}
\author{Daniel~Bothner}
\affiliation{Physikalisches Institut and Center for Collective Quantum Phenomena in LISA$^+$, Universit\"{a}t T\"{u}bingen, Auf der Morgenstelle 14, D-72076 T\"{u}bingen, Germany}
\author{Dieter~Koelle}
\affiliation{Physikalisches Institut and Center for Collective Quantum Phenomena in LISA$^+$, Universit\"{a}t T\"{u}bingen, Auf der Morgenstelle 14, D-72076 T\"{u}bingen, Germany}
\author{Reinhold~Kleiner}
\affiliation{Physikalisches Institut and Center for Collective Quantum Phenomena in LISA$^+$, Universit\"{a}t T\"{u}bingen, Auf der Morgenstelle 14, D-72076 T\"{u}bingen, Germany}
\author{Lapo~Bogani}
\affiliation{1. Physikalisches Institut, Universit\"{a}t Stuttgart, Pfaffenwaldring 57, D-70550 Stuttgart, Germany}
\author{Marc~Scheffler}
\affiliation{1. Physikalisches Institut, Universit\"{a}t Stuttgart, Pfaffenwaldring 57, D-70550 Stuttgart, Germany}
\author{Martin~Dressel}
\affiliation{1. Physikalisches Institut, Universit\"{a}t Stuttgart, Pfaffenwaldring 57, D-70550 Stuttgart, Germany}

\date{\today}

\begin{abstract}
We present non-conventional electron spin resonance (ESR) experiments based on microfabricated superconducting Nb thin film waveguides. A very broad frequency range, from 0.5 to 40 GHz, becomes accessible at low temperatures down to 1.6 K and in magnetic fields up to 1.4 T. This allows for an accurate inspection of the ESR absorption position in the frequency domain, in contrast to the more common observation as a function of magnetic field. We demonstrate the applicability of frequency-swept ESR on Cr$^{3+}$ atoms in ruby as well as on organic radicals of the Nitronyl-nitroxide family. Measurements between $1.6$ and $30$ K reveal a small frequency shift of the ESR and a resonance broadening below the critical temperature of Nb, which we both attribute to a modification of the magnetic field configuration due to the appearance of shielding supercurrents in the waveguide.
\end{abstract}


\maketitle

\begin{figure*}[tb]		%%%% FIG 1
	\centering
	\includegraphics[width=17cm]{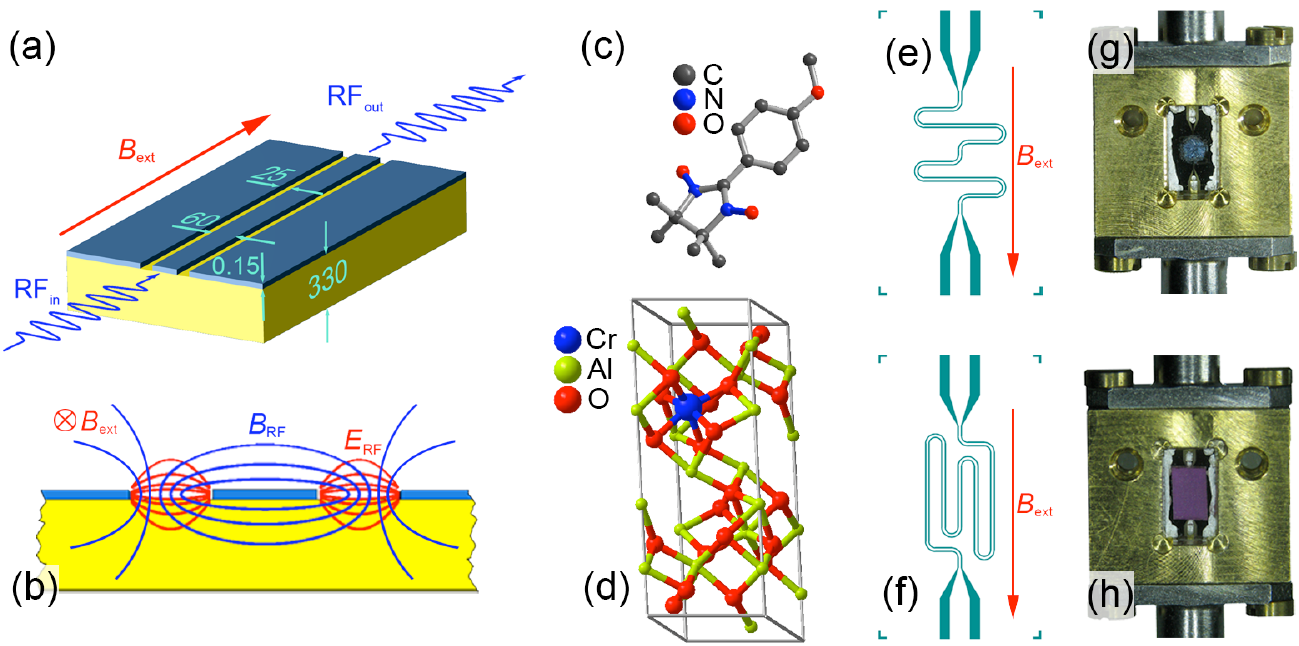}
	\caption{(color online) (a) Schematic layout of a coplanar waveguide and its orientation with respect to the external magnetic field in this experiment (numbers indicate lengths in $\micro\meter$). (b) Schematic cross-sectional view of the planar structure and sketched electric and magnetic field distribution of the microwave field. (c) Structure of NITPhOMe. The spin density is maximized at the N-O parts. (d) Unit cell of ruby. (e) and (f) Layout of the waveguide geometry used for the NITPhOMe and ruby experiments, respectively. (g) and (h) Photos of the mounted waveguides and samples.}
	\label{fig:Figure1}
\end{figure*}

To improve the performance of electron spin resonance (ESR) systems the main strategy was the use resonant of cavities with higher and higher fields and frequencies. As a result, most modern instrumentation can operate only at a single frequency or in an extremely narrow frequency range \cite{Poole97}. However, this can be a serious hindrance when a frequency-dependent effect is to be observed, as needed to probe a complete phase diagram of some material or to reliably assess the distance and orientation of spin labels. The latter problem, in particular, is increasingly important in the domain of biological ESR, as it is fundamental to understand the structural conformation and dynamics of biological systems. Until now the main strategy to overcome such problems was to measure the response of the system at a few discrete and widely-spaced frequencies \cite{Misra11}.\\
Another approach is to reduce the dimension of the ESR cavities to micrometer size while at the same time somewhat relaxing the resonance requirements \cite{Narkowicz05,Narkowicz08,Malissa12}. In this way small sample quantities can be probed over a wider frequency range. Setups that offer the possibility to sweep both the magnetic field and the radiation frequency have so far mostly been realized in the high frequency region (quasi-optical, from 50 to several $100~\giga\hertz$) \cite{Reijerse10,Gatteschi02,vanSlageren03}, while lower frequencies have been inspected using a coupled antenna approach \cite{Jang08}, tuneable cavities \cite{Schlegel10} or by placing the sample close to the center conductor of a coaxial line \cite{Rubinson89}.\\
In this letter we demonstrate a different approach which uses a microfabricated superconducting coplanar waveguide to generate the radio frequency (RF) field. The feasibility of such an approach was shown before by Schuster {\it et al.} focusing on high-cooperativity coupling of spin ensembles to superconducting cavities \cite{Schuster10}. Similar devices are also used to study ferromagnetic resonances of various materials \cite{Goglio99,Giesen05,Liu05,Harward11}. Using superconducting waveguides one can obtain higher RF fields (with same input power) and no ohmic heating of the sample, as will be shown later.
We show that it is possible to probe both the frequency and field dependence of ESR lines in a very wide frequency range from 0.5 to 40 GHz. We demonstrate the applicability of the method on a ruby single crystal [Al$_2$O$_3$:Cr$^{3+}$, see Fig. \ref{fig:Figure1}(d)], as well as on a Nitronyl-nitroxide radical, which was prepared from solution; 2-(4'-methoxyphenyl)-4,4,5,5-tetra-methylimidazoline-1-oxyl-3-oxide [NITPhOMe for short, see Fig. \ref{fig:Figure1}(c)]. NIT radicals are very clean and isotropic $S=\frac 12$ system that is widely used as spin labels in biology \cite{Jaeger08} and as a building block for molecular magnets \cite{Bogani11,Heintze13,Gentili05}. The results show that this technique is particularly suited to precisely determine the zero-field-splitting ($D$-values in the spin Hamiltonian) and the weight of higher order anisotropy terms of multilevel spin systems (which play an important role in e.g. rare earth coordination compounds \cite{Bogani11}).\\

Figure \ref{fig:Figure1}(a) shows a sketch of a coplanar waveguide. The signal-carrying center conductor line is flanked by two ground planes which act as the outer conductor/shielding. The devices are fabricated by sputtering a $150~\nano\meter$ thick niobium film on top of a $330~\micro\meter$ thick r-cut sapphire substrate and structuring the film with UV lithography (center conductor width: $60~\micro\meter$, center-ground plane separation: $25~\micro\meter$). The Nb films have a superconducting critical temperature $T_c\approx 9~\kelvin$ and a residual resistance ratio of $R_{300~\kelvin}/R_{10~\kelvin}\approx 6$. The resulting chip ($7\times 4\milli\meter^2$) is mounted into a gold-plated brass box and contacted with silver paste to sub-miniature A (SMA) stripline connectors (center conductor) and the box (ground planes). In case of the NITPhOMe sample, a small amount ($<0.5~\milli\gram$) of crystallites was dissolved in isopropyl and dropwise transferred to the waveguide. After evaporation of the solvent, the radicals were permanently attached to the waveguide structure (blue area in Fig. \ref{fig:Figure1}(g)). For the ruby sample, a polished brick-shaped crystal ($2.5\times 4\times 1~\milli\meter^3$) was put directly on the chip and fixed with vacuum grease (see Fig. \ref{fig:Figure1}(h)). This assembly is then cooled in a magnet cryostat. Unless noted otherwise, all data shown in this letter were acquired at $1.6~\kelvin$.

\begin{figure}[tb]		%%%% FIG 2
	\centering
	\includegraphics[width=8cm]{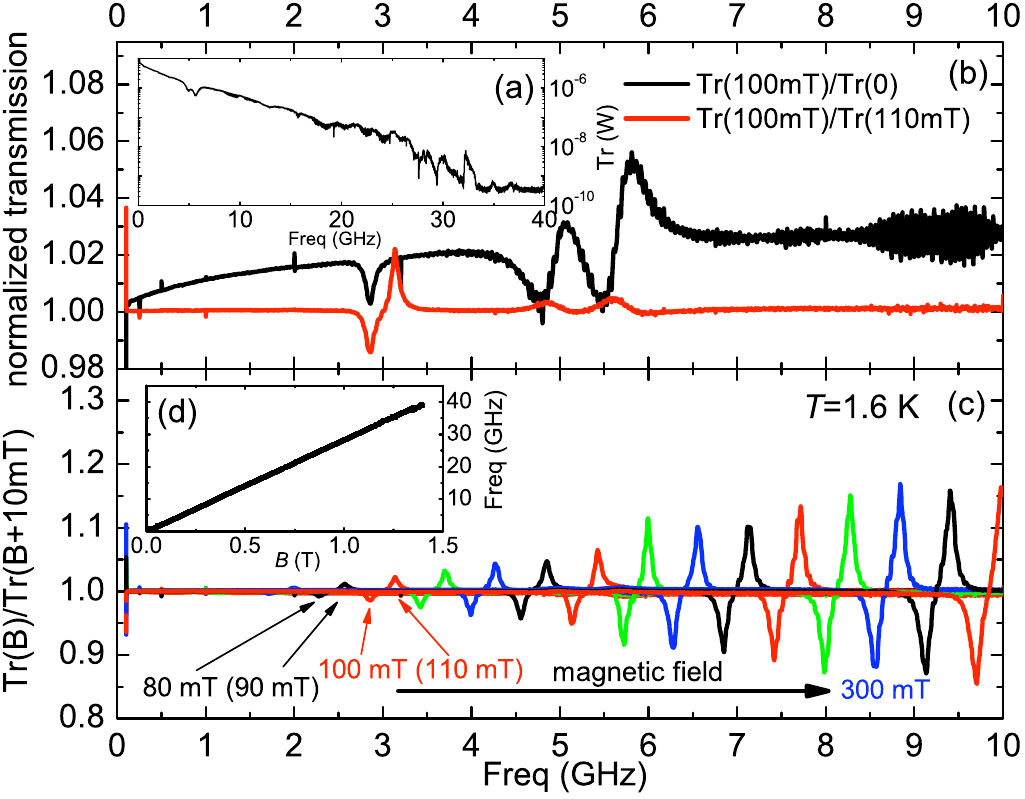}
	\caption{(color online) ESR spectra for NITPhOMe. (a) Broadband transmission spectra at zero field at $1.6~\kelvin$ for an input power of $-20~\deci\bel\milli$. (b) Example for the data analysis for the field-swept spectra. Black: Spectrum at $100~\milli\tesla$ normalized to zero field spectrum. Red/Gray: Spectrum at $100~\milli\tesla$ normalized to spectrum at $110~\milli\tesla$ (c) frequency-swept spectra for fields ranging from 20 to $340~\milli\tesla$ at $20~\milli\tesla$ steps (very low field absorptions not visible at this scale). (d) Position of the absorption peak minimum vs. magnetic field for all measured spectra.}
	\label{fig:Figure2}
\end{figure}

For all the presented experiments the external magnetic field is oriented parallel to the film (see Figure \ref{fig:Figure1}(a))\footnote{The experimental insert is designed for a film orientation parallel to the magnetic field. At this moment, we cannot quantify the misalignment of the field, but we estimate a tilting out of plane of less than 2\degree.}. Note that only for an orientation as shown in Figure \ref{fig:Figure1}(a) the magnetic field component of the microwave signal ($B_{\rm RF}$) is perpendicular to the static external field ($B_{\rm ext}$) and thus fulfills the requirement to induce spin flips between the Zeeman-split levels (see also Figure \ref{fig:Figure1}(b)). For the designs used, the waveguide meanders over the chip and only parts of it are oriented in the correct way (see Fig. \ref{fig:Figure1}(e),(f)). While for the NITPhOMe sample this portion is only $120~\micro\meter$, it is much longer ($8~\milli\meter$) for the measurements on the ruby sample.\\
In both cases a continuous wave microwave signal (input power $P_{\rm in}=-20~\deci\bel\milli$) is fed via coaxial cables to the cold waveguide with the sample, and the transmitted power is recorded with a power meter (NITPhOMe) or with a vector network analyzer (ruby).
For the frequency-swept spectra the transmitted power is recorded as a function of frequency at a static external field $B_{\rm ext}$. Results for NITPhOMe are shown in Figure \ref{fig:Figure2}(a). The transmission decreases with increasing frequency due to the frequency-dependent attenuation of the coaxial cables and the coplanar line.\\
To identify the signal arising from spin transitions in the sample, the spectra taken at finite external magnetic field $B_{\rm ext}$ have to be normalized. The simplest approach is using the zero field spectrum for normalization of all other spectra (see Figure \ref{fig:Figure2}(b)). This however turned out to be problematic, since in addition to a change in the overall transmission through the superconducting waveguide due to losses resulting from movement of vortices \cite{Song09,Bothner12} there are field-dependent features appearing around certain frequencies. These features appear in the normalized spectra and can obscure the ESR signal (see upper curve in Figure \ref{fig:Figure2}(b)). These features (around $5~\giga\hertz$) are box modes and standing waves on the waveguide between the connectors or on the ground planes. They shift in magnetic field due to the change of the high frequency transport properties of the superconducting film. These spurious modes change slightly for different waveguides or sample holder boxes. It is therefore not trivial to compensate for them.\\
\begin{figure}[tb]		%%%% FIG 3
	\centering
	\includegraphics[width=8cm]{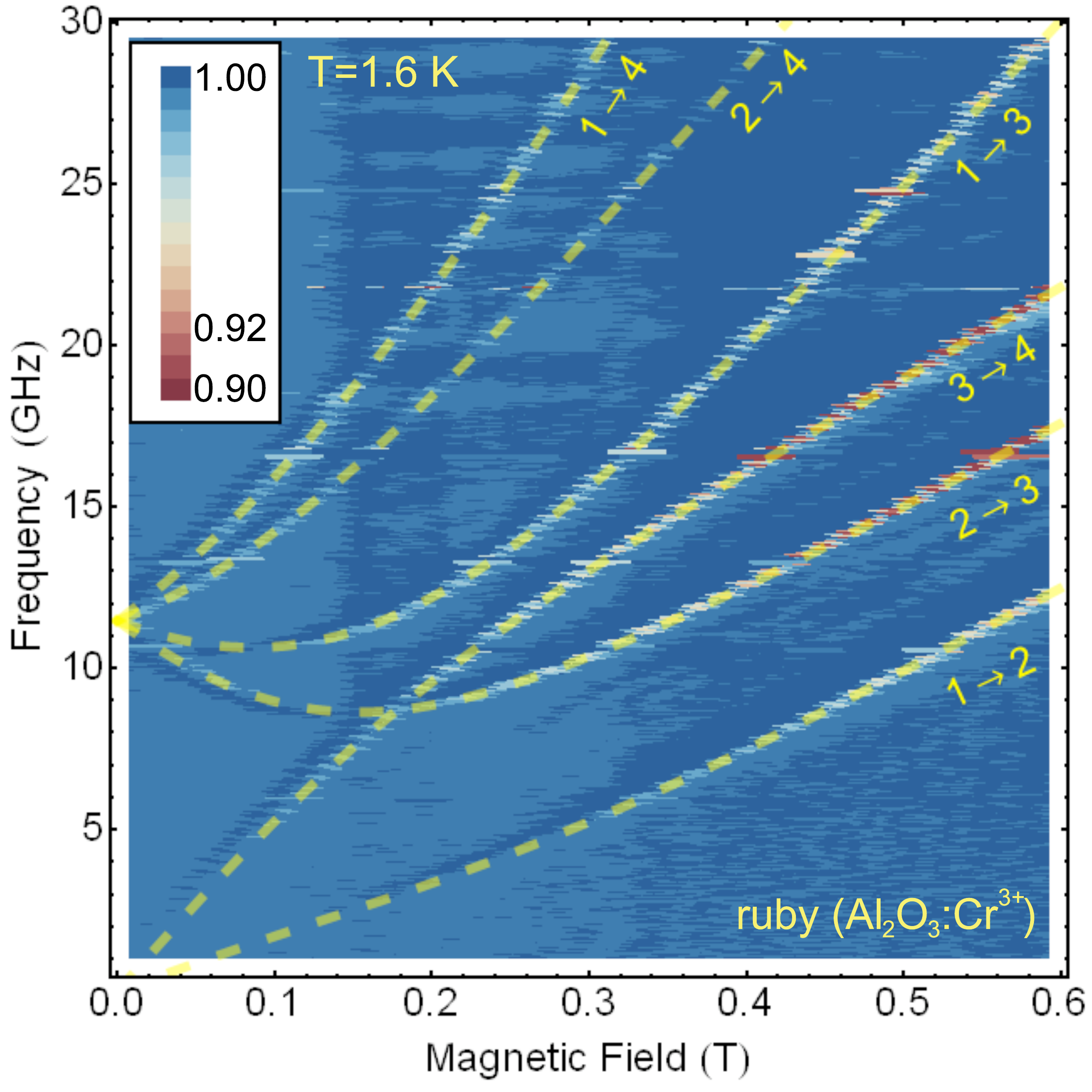}
	\caption{(color online) Normalized frequency-swept ESR spectra for ruby. Lighter and red areas show the ESR absorption positions and dashed lines show a fit to the transitions between the eigenfunctions of the Cr$^{3+}$ spin Hamiltonian. The different spin states are labeled as 1: $m_s=-\frac 32$, 2: $m_s=-\frac 12$, 3: $m_s=+\frac 12$ and 4: $m_s=+\frac 32$}
	\label{fig:Figure3}
\end{figure}
However, since both the increase of attenuation of the superconducting waveguide as well as the variation of box mode absorption evolve only slowly with magnetic field, one can use a spectrum taken at a slightly different field for normalization. Here we found that spectra taken at $10~\milli\tesla$ higher fields result in a straight baseline at unity and a rather good suppression of the evolving parasitic features (see lower line in Figure \ref{fig:Figure2}(b)). A selection of spectra normalized in this manner is shown in Figure \ref{fig:Figure2}(c) for fields from $20$ to $340~\milli\tesla$ in $20~\milli\tesla$ steps for a frequency range from $100~\mega\hertz$ to $10~\giga\hertz$ (in fact the absorption for every $10~\milli\tesla$ can be seen since the peaks above unity belong to the spectra used for normalization). The signal intensity increases with increasing external field due to the increasing thermal population difference of the Zeeman split energy levels.\\
As can be seen from Figure \ref{fig:Figure2}(a) the transmission steadily decreases with increasing frequency up to about $33~\giga\hertz$. Due to the connectors, which are not recommended for frequencies above $25~\giga\hertz$, the measured transmitted power for higher frequencies is close to the detection limit of the power meter and the spectra get very noisy. Despite the high noise level, the absorption peaks can still be identified and their position is plotted vs. the external magnetic field in Figure \ref{fig:Figure2}(d). From the slope of that curve ($h \nu=g \mu_B B_{\rm ext}$) a $g$-value of $g=2.019\pm 0.001$ was extracted.\\
We also performed frequency-swept ESR on a ruby sample of arbitrary crystallographic orientation for a frequency range from 1 to $30~\giga\hertz$ and fields up to $0.6~\tesla$ in $7.5~\milli\tesla$ steps. The resulting normalized spectra (Tr($B$)/Tr($B+0.015~\tesla$)) are shown as a color map plot in Fig. \ref{fig:Figure3} for a range from $0.9$ to $1.0$ (to not show the peaks of the reference spectra). All possible transitions between the four states of the zero-field-split quartet can be seen over almost the complete magnetic field and frequency range. The lighter regions denote the positions of ESR absorptions and dashed lines are transitions calculated from the Cr$^{3+}$ spin Hamiltonian for an orientation of $B_{\rm ext}$ at $78\degree$ to the $c$-axis of the crystal \cite{Weber59,SchulzDuBois59}. Since every frequency-swept spectrum can be used to fit unknown parameters of the spin Hamiltonian, this method could be used to determine the weight of higher order anisotropy terms \cite{Bernot09} for more complex (low symmetry) materials.\\
In addition to the frequency-swept spectra, we also performed magnetic-field-swept ESR. Several selected spectra are shown in Figure \ref{fig:Figure4} for NITPhOMe. Here, we record the transmitted power at fixed frequencies as a function of applied external field. The off-resonance transmission decreases with increasing field/frequency due to the frequency-dependent attenuation mentioned above. Figure \ref{fig:Figure4} (b) shows a set of field-swept spectra taken at the upper edge of the accessible frequency range, demonstrating that ESR can be detected for frequencies as high as $40~\giga\hertz$.\\

Considering the sensitivity of the technique, the lowest accessible frequency can be used to estimate the minimum number of required spins. For the two level spin $\frac 12$ NITPhOMe sample the ESR absorption signal could be observed for transition frequencies down to about $500~\mega\hertz$ (at $1.6~\kelvin$). The total number of spins on the chip was determined by the total mass of the transferred radicals to be in the order of $10^{18}$. Since only a small section of the waveguide ($120~\micro\meter$ long part seen in the center of Fig. \ref{fig:Figure1}e) is oriented so that $B_{\rm ext}\perp B_{\rm RF}$ and since the microwave magnetic field decreases very fast with increasing distance from the gaps between center conductor and ground planes \cite{Schuster10} only a small portion of about $10^{15}$ spins are responsible for the observed ESR absorption. This is the estimated minimum number of spins for the low frequency limit. For higher transition frequencies the minimum number of spins decreases, so that e.g. for $30~\giga\hertz$ only about $2.5\cdot 10^{13}$ spins are required. At this point, this sensitivity cannot compete with typical sensitivities achieved with resonant cavity systems. This does not come as a surprise because our main focus is the broadband nature of the approach, which intrinsically leads to lower sensitivity than resonant techniques. But we expect that our approach can easily reach much better sensitivity by improving the detection mechanism, e.g. by lock-in detection in combination with modulation of the static field.\\

\begin{figure}[tb]		%%%% FIG 4
	\centering
	\includegraphics[width=8cm]{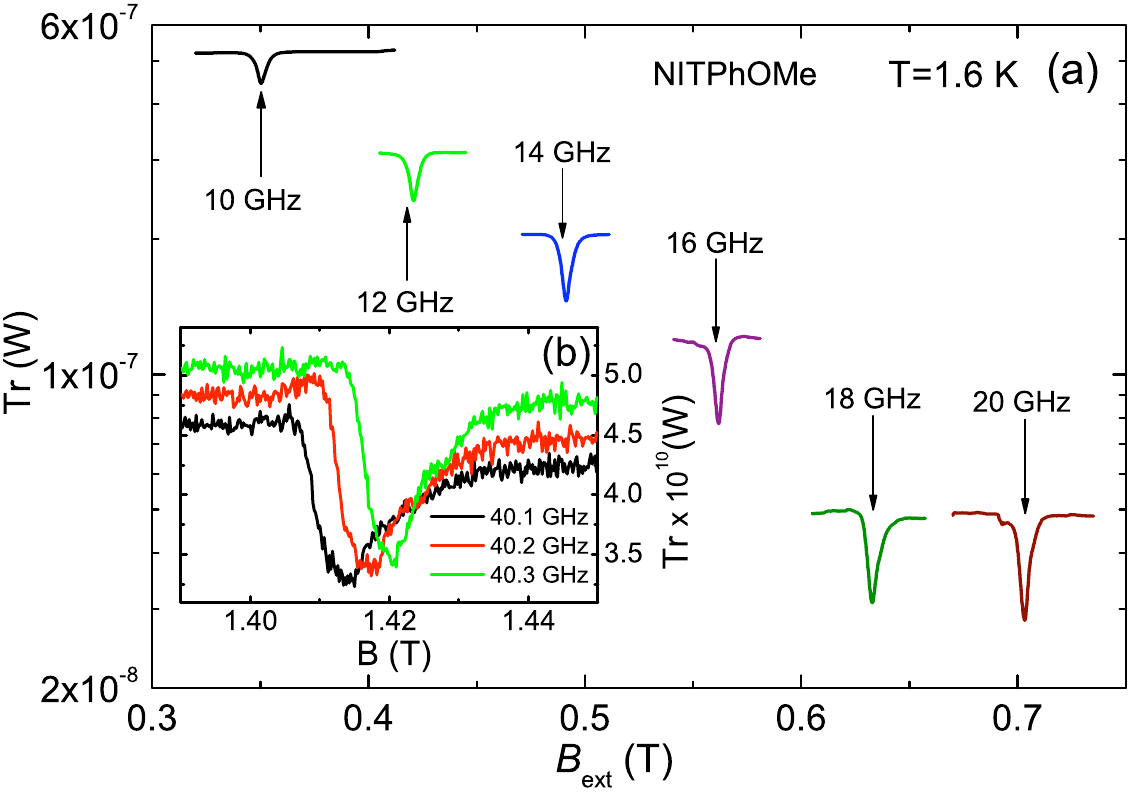}
	\caption{(color online) (a) Field-swept spectra for various frequencies for an input power of $-20~\deci\bel\milli$. The absolute value of the transmitted power off-resonance is predominantly determined by the frequency-dependent damping of the coaxial lines. Inset (b): Field swept spectra  around $40~\giga\hertz$.}
	\label{fig:Figure4}
\end{figure}
\begin{figure}[b]		%%%% FIG 5
	\centering
	\includegraphics[width=8cm]{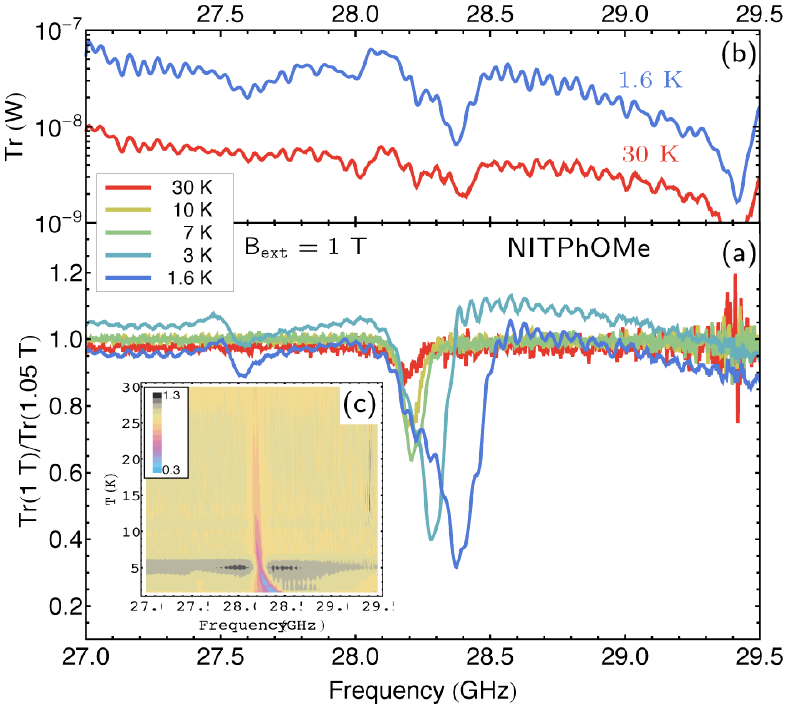}
	\caption{(color online) (a) Frequency-swept spectra for several selected temperatures. The absorption is already visible for temperatures above $T_c$ of Nb. Upon cooling through the superconducting transition, the absorption peak frequency shifts to higher frequencies. (b) Transmitted power at $1~\tesla$ for $1.6~\kelvin$ and $30~\kelvin$. (c) All measured spectra shown in a contour map. The transition into the superconducting state around $5~\kelvin$ as well as the shift of the ESR absorption frequency for temperatures below $T_c$ is clearly visible.}
	\label{fig:Figure5}
\end{figure}
Although the high frequency losses of the superconducting waveguide are minimized when working at temperatures below $T_c$, we were also able to detect the ESR signal of NITPhOMe at temperatures well above $T_c$. The temperature dependence of a selection of frequency-swept spectra for $B_{\rm ext}=1~\tesla$ (spectra were normalized with those at $B_{\rm ext}=1.05~\tesla$) is shown in Fig. \ref{fig:Figure5} (a). The transmitted power in the superconducting state (for this waveguide) is about one order of magnitude higher than that in the normal state as shown in \ref{fig:Figure5} (b). With constant input power one achieves higher currents and therefore higher $B_{\rm RF}$ fields in the superconducting state. This leads to a larger volume within which the microwave fields are strong enough to induce ESR absorptions and therefore increases the number of spins that contribute to the ESR signal.\\\\
Upon cooling down from $30~\kelvin$, we observe an increase of the peak amplitude due to the increase of imbalance of thermal occupation. Reaching the critical temperature of the Nb film (for these fields at around $5~\kelvin$) and cooling further, two effects can be observed. (i) While crossing the transition temperature, the baseline deviates from unity due to different transport properties of the metallic film for the spectra recorded at $B_{\rm ext}=1~\tesla$ compared to the reference spectrum recorded at $B_{\rm ext}=1.05~\tesla$. (ii) Below $T_c$ the peak frequency of the ESR signal shifts to slightly higher frequencies and the absorption dip gets broadened. We attribute this effect to a modification of the local dc magnetic field $B_{\rm ext}$ by shielding currents in the superconductor. Directly at the surface of the Nb planes -- and hence also in the gaps between center conductor and ground planes -- the magnetic field is enhanced as compared to the normal conducting state, which shifts the Zeeman splitting of spins sitting directly at the surface to higher values. With increasing distance from the Nb surface the field values approach $B_{\textrm{ext}}$ and due to this field inhomogeneity we not only get a shift but also a broadening of splitting energies and of the ESR absorption dip. However, the effect of the modified field profile on the shift of the resonance frequency in our experiments is quite small (about 1\%) but nevertheless has to be considered when it comes to the precise determination of ESR frequencies with superconducting waveguides. Although this effect is disadvantageous for the detection of ESR it could be used to quantitatively study the local field variations at the border or at structural gaps of superconductors. For high-precision measurements, an ESR standard such as diphenyl picrylhydrazyl (DPPH) can be used to account for the change in the applied magnetic field due to expelled flux of the Nb film.\\\\
In conclusion, we have shown that microfabricated superconducting structures can be used for broadband ESR spectroscopy over a wide range of frequencies and magnetic fields. The temperature-dependent results show a good performance of our system even above the critical temperature and suggest the use of non-superconducting materials to probe ESR over an even wider range of temperatures. Due to the compact design, the device could easily be used down to $1.6~\kelvin$ and provides the possibility to perform ESR experiments in a dilution refrigerator, reaching the mK regime. A complete frequency-field characterization is now possible (see Fig. \ref{fig:Figure3}) and the technique thus offers the possibility to characterize a variety of materials with high precision. In particular it is suited to determine the orientation of anisotropy axes \cite{Bernot09}, large zero-field-splittings \cite{Bogani11} and will provide access to higher-order anisotropy terms of high-spin systems \cite{Bogani11,Bernot09}. It is thus expected to be particularly useful for the characterization of molecular materials and rare-earth based magnets \cite{Bogani11}. The combination of a wide frequency and field range will also be useful for the spectroscopic determination of level mixings (avoided level crossings), which are expected in many magnetic molecular clusters. Eventually it is important to understand that the technique can readily be exported to the pulsed domain. The application to Nitronyl-nitroxide spin-labels shows that the technique can be applied to provide the accurate, frequency-domain information needed for the structural characterization of biological structures \cite{Jaeger08}.\\\\
This work was supported by the Deutsche Forschungsgemeinschaft, including SFB/TRR 21 and SPP1601 and the Humboldt Stiftung via the Sofja Kovalevskaja prize. DB gratefully acknowledges support by Evangelisches Studienwerk e.V. Villigst.

\end{document}